\documentclass[lettersize,journal]{IEEEtran}
\usepackage{cite}
\usepackage{amsmath,amssymb,amsfonts}
\usepackage{graphicx}
\usepackage{textcomp}
\usepackage{dsfont}
\usepackage{algpseudocode}
\usepackage{algorithm}
\usepackage{multirow, tabularx}
\usepackage{array}
\usepackage{subcaption}
\usepackage[caption=false,font=normalsize,labelfont=sf,textfont=sf]{subfig}

\begin{document}


\title{EEG\_RL-Net: Enhancing EEG MI Classification through Reinforcement Learning-Optimised Graph Neural Networks}

\author{Htoo Wai Aung, Jiao Jiao Li, Yang An, and Steven W. Su*, \IEEEmembership{Senior Member, IEEE} \thanks{Htoo Wai Aung is the first author of this paper, and he is with the School of Biomedical Engineering, Faculty of Engineering and IT, University of Technology Sydney, NSW 2007, Australia (e-mail: htoowai.aung@student.uts.edu.au).
} 

\thanks{Jiao Jiao Li is with the School of Biomedical Engineering, Faculty of Engineering and IT, University of Technology Sydney, NSW 2007, Australia (e-mail: jiaojiao.li@uts.edu.au).}

\thanks{Yang An is with the Faculty of Engineering and IT, University of Technology Sydney, NSW 2007, Australia (e-mail: yang.an-1@student.uts.edu.au).}

\thanks{Steven W. Su is with the Faculty of Engineering and IT, University of Technology Sydney, NSW 2007, Australia ( e-mail: steven.su@uts.edu.au).}
\thanks{*Corresponding author.}
}

\markboth{Journal of \LaTeX\ Class Files,~Vol.~xx, No.~x, Apr~2024}%
{Shell \MakeLowercase{\textit{et al.}}: A Sample Article Using IEEEtran.cls for IEEE Journals}

\maketitle

\begin{abstract}

Brain-Computer Interfaces (BCIs) rely on accurately decoding electroencephalography (EEG) motor imagery (MI) signals for effective device control. Graph Neural Networks (GNNs) outperform Convolutional Neural Networks (CNNs) in this regard, by leveraging the spatial relationships between EEG electrodes through adjacency matrices. The EEG\_GLT-Net framework, featuring the state-of-the-art EEG\_GLT adjacency matrix method, has notably enhanced EEG MI signal classification, evidenced by an average accuracy of 83.95\% across 20 subjects on the PhysioNet dataset. This significantly exceeds the 76.10\% accuracy rate achieved using the Pearson Correlation Coefficient (PCC) method within the same framework. 

In this research, we advance the field by applying a Reinforcement Learning (RL) approach to the classification of EEG MI signals. Our innovative method empowers the RL agent, enabling not only the classification of EEG MI data points with higher accuracy, but effective identification of EEG MI data points that are less distinct. We present the EEG\_RL-Net, an enhancement of the EEG\_GLT-Net framework, which incorporates the trained EEG\_GCN Block from EEG\_GLT-Net at an adjacency matrix density of 13.39\% alongside the RL-centric Dueling Deep Q Network (Dueling DQN) block. The EEG\_RL-Net model showcases exceptional classification performance, achieving an unprecedented average accuracy of 96.40\% across 20 subjects within 25 milliseconds. This model illustrates the transformative effect of the RL in EEG MI time point classification.

\end{abstract}

\begin{IEEEkeywords}
Brain-Computer Interfaces (BCIs), Electroencephalography Motor Imagery (EEG MI), Spectral Graph Convolutional Neural Networks (GCNs), Reinforcement Learning (RL), Dueling Deep Q Network (Dueling DQN)
\end{IEEEkeywords}

\section{Introduction}
\label{sec:introduction}

\IEEEPARstart{B}{rain-Computer Interfaces} establish a connection between the brain and external control devices. Originally developed to assist individuals with motor impairments \cite{WOLPAW2002767}, BCIs translate brain signals acquired through measurements such as electrocorticography (ECoG) and electroencephalogram (EEG) into actionable commands for electronic control devices including wheelchairs and exoskeleton robots. Although ECoG offers superior signal quality over EEG, its application in BCIs is limited due to invasive route of acquisition, requiring the placement of electrodes directly on the cerebral cortex \cite{lebedev2006brain}. Meanwhile, EEG is a much more accessible and hence popular signal acquisition method as it involves non-invasive placement of electrodes on the scalp. EEG is widely used to record various types of brain signals, from spontaneous and stimulus-evoked signals to event-related potentials \cite{schomer2012niedermeyer}. Its clinically relevant applications extend to dementia classification \cite{cao2024dementia}, depression state assessment \cite{wang2024eeg}, seizure detection \cite{cappelletti2024learning}, and the classification of cognitive and motor tasks \cite{ding2023lggnet}, including motor imagery (MI) tasks \cite{hou2022gcns, abbasi2024novel, wang2024eeg}. 

MI involves the mental simulation of motor actions, such as movements of the hands, feet, or tongue, without performing the physical movements \cite{hubbard2019eeg, mcfarland2000mu}. This technique is crucial in neuroscience and rehabilitation, with real-world relevance especially for individuals with motor impairments, such as stroke survivors. Through integration with an external control device, MI enables the physically impaired to perform daily activities that are not otherwise possible, leading to potentially life-changing benefits by improving quality of life and reducing the level of chronic care. By integrating MI and BCIs, EEG based MI signals can be decoded and used to control external devices, enabling real-time feedback and facilitating patient-intended movements through accurate signal interpretation \cite{biasiucci2018brain}.

Deep learning, a subset of machine learning, utilises multiple layers of neural networks to process a variety of data forms. Convolutional Neural Networks (CNNs), which mimic natural image recognition in the human visual system, are part of the deep learning family and excel in computer vision tasks \cite{farabet2012learning, lecun1998gradient, lecun2010convolutional}. However, their application is restricted to Euclidean data, such as 1-dimensional sequences and 2-dimensional grids \cite{lecun2010convolutional}. CNNs struggle with non-Euclidean data, failing to accurately capture the intrinsic structure and connectivity of the data.

Graph Convolutional Networks (GCNs) have been developed to perform convolutional operations on graphs, which can handle non-Euclidean data due to incorporating topological relationships during convolution. GCNs can represent complex structures and variations in these structures, which may be heterogeneous or homogeneous, weighted or unweighted, signed or unsigned \cite{zhang2020deep}. They support various types of graph analyses, including node-level, edge-level, and graph-level tasks \cite{wu2020comprehensive, zhang2020deep}. GCNs are particularly effective at classifying EEG signals as a graph-level task \cite{hou2022gcns, aung2024eeggltnet}. For this application, EEG signal readings from each channel are treated as node attributes, and the relationships between EEG electrodes are represented by an adjacency matrix, hence surpassing the capabilities of traditional CNNs.

There are two primary categories of GCNs: spatial \cite{hamilton2017inductive, monti2017geometric, niepert2016learning, gao2018large} and spectral methods \cite{bruna2013spectral, defferrard2016convolutional, levie2018cayleynets}. Some challenges are encountered with the spatial method \cite{shuman2013emerging, bao2022linking} especially in matching local neighbourhoods. Both time domain and frequency domain features can be extracted from EEG signals to perform GCN operations \cite{zeng2020hierarchy, meng2022electrical, cao2024dementia, li2023sequential}. Frequency domain features include Power Spectral Density (PSD) and Power Ratio (PR) for various bands, such as $\delta$ (0.5-4Hz), $\theta$ (4-8Hz), $\alpha$ (8-13Hz), $\beta$ (13-30Hz), and $\gamma$ (30-110Hz) within specified time windows. Time domain features, such as Root Mean Square (RMS), skewness, minmax, variance, number of zero crosses, Hurst Exponent, Petrosian fractal, and Higuchi, are also extracted for GCN operations during specific time windows. These features are integral to window-based GCN methods.

In the GCNs-Net \cite{hou2022gcns}, individual time point signals at each channel are treated as distinct features. This method is designed for real-time EEG MI signal classification, focusing on $\frac{1}{160}s$ time point signals. The constructing of an effective adjacency matrix is crucial for GCN operations, and different methods have been explored in various studies, including: Geodesic method, which relies on geodesic distances between EEG channels \cite{abbasi2024novel, zhang2022recognizing, jia2022efficient, wagh2020eeg}; using Pearson Coefficient Correlation (PCC) to evaluate interchannel correlations \cite{khaleghi2023developing, ma2023double, bao2022linking, meng2022electrical, hou2022gcns} ; and experimenting with a trainable matrix approach \cite{bao2022linking, song2018eeg}.

In the EEG\_GLT-Net \cite{aung2024eeggltnet}, a sophisticated algorithm known as the EEG Graph Lottery Ticket (EEG\_GLT) is used to optimise the adjacency matrix by exploring various density levels, inspired from the unified GNN sparsification technique (UGS) \cite{chen2021unified}. This method represents the current state-of-the-art in adjacency matrix construction, significantly enhancing accuracy, F1 score, and computational efficiency on the EEG MI PhysioNet dataset \cite{physionet_dataset} compared to the PCC and Geodesic methods. However, despite the overall superiority of this method, it remains challenging to classify the EEG MI time points remains challenging for some subjects due to signal ambiguity among different MI tasks at specific time points. Consequently, supervised learning on these subjects involves training that forces classification of all time points.

Reinforcement Learning (RL), another subset of machine learning, enables an RL agent to learn sequential decision-making in dynamic environments to maximise cumulative rewards \cite{sutton2018reinforcement}. RL has been primarily applied in robotics and autonomous systems, which require complex sequential decision-making. Deep RL principles have been applied to optimise feature selection for the Classification with Costly Features (CwCF) problem \cite{janisch2019classification}, across various public UCI datasets \cite{Dua:2019} including miniboone, forest, cifar, wine, and mnist. Others \cite{song2018deep} have trained an RL agent to minimise feature extraction costs in classifying electromyography (EMG) signals from UCI datasets \cite{Dua:2019}, although this reduction in features compromised accuracy.

In this paper, we introduce EEG\_RL-Net as a new algorithm, with more advanced capability than existing methods for classifying EEG MI time point signals by combining GNNs and RL. Initially, optimal graph features of EEG MI time point signals are extracted using the best weights and adjacency matrix from an EEG\_GCN block, refined to 13.39\% density using the EEG\_GLT algorithm. Subsequently, the RL agent makes sequential decisions within an episode of pre-defined horizon length to accurately classify the EEG MI signals. The main contributions of this study are:

\begin{itemize}
    \item \textbf{EEG\_RL-Net:} A new approach for classifying EEG MI time point signals, using a trained RL agent that determines whether to classify or skip each time point based on GNN features. This method greatly enhances performance accuracy by achieving classification as swiftly as possible within predefined episode lengths.

    \item \textbf{Optimal Reward and Max Episode Length Setting:} We evaluated the accuracy and classification speed under various reward settings and maximum episode lengths for each subject, identifying the optimal combinations for simultaneously achieving high accuracy and efficiency.

    \item \textbf{Performance Validation:} We evaluated the performance of each subject under optimal settings against the state-of-the-art EEG\_GLT-Net with $m_{g\_GLT}$ matrix and PCC adjacency matrix. Our results showed significant enhancement of accuracy and efficiency on the PhysioNet dataset.
\end{itemize}

\section{Methodology}
\subsection{Overview}

\begin{figure*}
    \centering
    \includegraphics[width=0.85\textwidth]{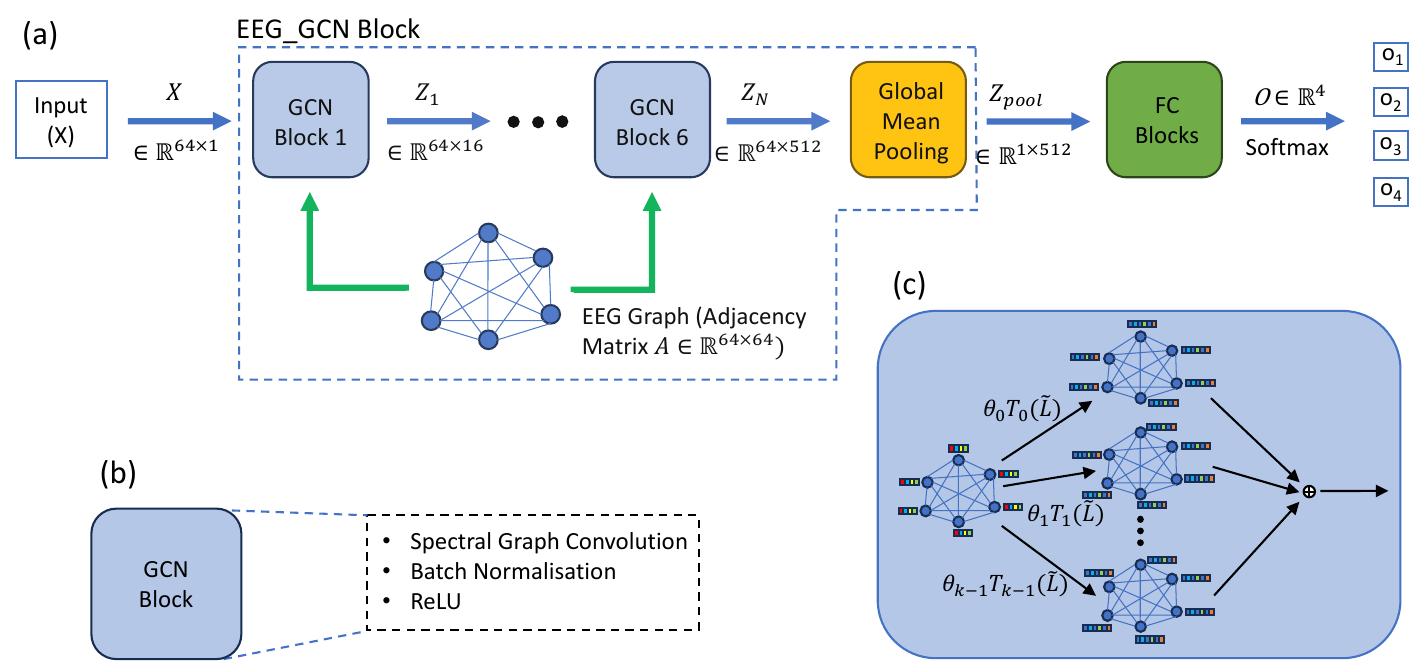}
    \caption{EEG\_GLT-Net model \cite{aung2024eeggltnet}: (a) Overall architecture (classifying EEG MI of one time point $\frac{1}{160}s$ of signals from 64 EEG electrodes), (b) Components inside the spectral graph convolution block, (c) Chebyshev spectral graph convolution}
    \label{fig: eeg_glt_net_overview}
\end{figure*}

This project is divided into two distinct parts. The first phase focused on training the EEG\_GLT-Net model, as illustrated in Figure~\ref{fig: eeg_glt_net_overview}, to identify the optimal adjacency matrices and spectral GNN weights across different adjacency matrix density levels, employing Algorithm~\ref{alg: EEG_GLT algorithm}. This phase of training spanned from t = 1s to t = 3s. Subsequently, the optimal adjacency matrix and spectral GNN weights, determined at the minimal adjacency matrix density level of 13.39\%, were selected for the purpose of extracting graph features.

In the project's second phase, the Multilayer Perceptron (MLP) block within the EEG\_GLT-Net was removed, and in its place, the RL (Reinforcement Learning) block was integrated, resulting in the formation of the EEG\_RL-Net, as depicted in Figure~\ref{fig: eeg_glt_rl_overview}. The pre-trained optimal weights of the EEG\_GCN block, such as adjacency matrix and spectral GNN, determined at the lowest adjacency matrix density of 13.39\%, were then transferred to the EEG\_GCN component of the EEG\_RL-Net architecture. During this phase, all time points from $t=0s$ to $t=4s$ were utilised, with these points organised into groups spanning a horizon of 20 states, where each point represented a single state. Within each episode's horizon, the RL agent performed action at every state, based on the graph features generated by the GNN segment. These actions involved classifying the state as belonging to Task 1 through to Task 4, or skipping to the next state (Task 0) if the agent determined that it was not yet prepared to classify.

\subsection{Dataset Description and Pre-processing}
Following the approach of papers \cite{hou2022gcns} and \cite{aung2024eeggltnet}, this study employed the PhysioNet EEG MI dataset \cite{physionet_dataset}, which comprises EEG recordings from 109 subjects acquired using the international 10-10 system with 64 EEG channels. The dataset is structured around four distinct EEG MI tasks, which involve the subject imagining the actions of:
\begin{itemize}
    \item Task 1: Opening and closing the left fist.
    \item Task 2: Opening and closing the right fist.
    \item Task 3: Opening and closing both fists simultaneously.
    \item Task 4: Opening and closing both feet.
\end{itemize}

Each participant completed 84 trials, divided into 3 runs with 7 trials per run for each task type. The duration of each trial's recording was 4 seconds, sampled at 160Hz. In our study analyses were specifically conducted on a subset of 20 subjects, labelled $S_1$ to $S_{20}$. Initially, the raw signals were processed solely through a notch filter at the 50Hz power line frequency to eliminate electrical interference, deliberately avoiding other common filtering or denoising techniques to preserve data integrity. Signals from all 64 channels were utilised, with each channel treated as a node and the signal at each time point considered as the node's feature. Additionally, the signals at each channel were normalised to achieve a mean ($\mu$) of 0 and a standard deviation ($\sigma$) of 1.

\subsection{Graph Feature Extraction}
\subsubsection{Graph Representation}
In a directed graph, $G=\{V, E\}$ where $V= \{v_1, v_2, ..., v_N\}$ represents the set of nodes and $|E|$ signifies the total number of edges connecting these nodes. The structure of the graph can be illustrated using an adjacency matrix $A\in \mathbb{R}^{N\times N}$. Every node within the graph is associated with $F_N$ features, and the matrix encapsulating these node features is expressed as $X=\in \mathbb{R}^{N\times F_N}$. A combinatorial Laplacian matrix, denoted as $L$ , is derived through Equation~(\ref{eqn: L_norm}). This involves the use of the degree matrix of $A$, symbolised as $D$, which is calculated using $D_{ii}=\sum_{j=1}^{N}A_{ij}$.

\begin{equation}
\label{eqn: L_norm}
L=I_{N} - D^{-1/2} A D^{-1/2}
\end{equation}

\subsubsection{Spectral Graph Filtering}
The eigenvectors of the graph Laplacian matrix can be expressed in the Fourier mode as $\{u_l\}_{l=0}^{N-1}\in \mathbb{R^N}$, with the Fourier basis $U=[u_0, ..., u_{N-1}]\in \mathbb{R}^{N\times N}$. The corresponding eigenvalues, denoted as $\{\lambda_l\}_{l=0}^{N-1}\in \mathbb{R}$, represent the graph Fourier frequencies, and the diagonal matrix containing these Fourier frequencies, $\Lambda$, is defined as $\Lambda=diag[\lambda_0, ..., \lambda_{N-1}] \in \mathbb{R}^{N\times N}$. A signal $x$ can undergo a graph Fourier transform to become $\hat{x}=U^Tx$, and the inverse Fourier transform is obtained with $x=U\hat{x}$. The convolution operation on the graph $G$ is defined as:

\begin{equation}
\label{eqn: graph conv 1}
x *_{G} g = U((U^T x) \odot (U^T g))
\end{equation}

\noindent where $g\in \mathbb{R}^{N}$ denotes a convolutional filter. With $g_{\theta}(\Lambda)=diag(\theta)$, where $\theta \in \mathbb{R}^N$ symbolises the vector of Fourier coefficients, the graph convolution of the signal $x$ is executed as:

\begin{equation}
\label{eqn: graph conv 2}
x *_G g_\theta = U g_\theta (\Lambda) U^T x
\end{equation}

Given the non-parametric and non-localised nature of the $g_{\theta}$ filter, its computational demand is excessively high. Utilising the Chebyshev graph convolution technique, the computational complexity is reduced from $O(N^2)$ to $O(KN)$. The $g_{\theta}$ approximation, up to the $K^{th}$ order within the Chebyshev polynomial framework, is facilitated using Equation~\ref{eqn: graph cheby}. The normalisation of the $\Lambda$ can be achieved using Equation~\ref{eqn: A_lambda norm}. The term $\theta_k$ denotes the coefficients of the Chebyshev polynomial, and $T_k(\hat{\Lambda})$ is derived using Equation~\ref{eqn: graph cheby norm}.

\begin{equation}
\label{eqn: graph cheby}
g_\theta (\Lambda) =  \sum_{k=0}^{K-1}\theta_k T_k (\hat{\Lambda})
\end{equation}

\begin{equation}
\label{eqn: A_lambda norm}
\hat{\Lambda}=\frac{2\Lambda}{\Lambda_{max}} - I_N
\end{equation}

\begin{equation}
\label{eqn: graph cheby norm}
\{T_0(\hat{\Lambda})=1,  T_1=(\hat{\Lambda}), T_k(\hat{\Lambda})=2\hat{\Lambda}T_{k-1}(\hat{\Lambda})-T_{k-2}(\hat{\Lambda})\}
\end{equation}

Ultimately, the graph convolution operation on the signal $x$ is executed as shown in Equation~\ref{eqn: graph cheby 2}, utilising the normalised Laplacian matrix, $\widetilde{L}$ which is calculated through Equation~\ref{eqn: lambda norm}.

\begin{equation}
\label{eqn: graph cheby 2}
x *_G g_\theta = U \sum^{K-1}_{k=0}\theta_k T_k(\hat{\Lambda})U^T x = \sum^{K-1}_{k=0} \theta_k T_k (\widetilde{L})x
\end{equation}

\begin{equation}
\label{eqn: lambda norm}
\widetilde{L}=\frac{2L}{\lambda_{max}}-I_N
\end{equation}

\subsubsection{Training EEG\_GLT-Net}
In the EEG\_GLT-Net study \cite{aung2024eeggltnet}, the classification of EEG MI signals, $X$ is facilitated through a forward pass using the Spectral GNN function, denoted as $f(., \Theta)$, with a given graph $G=\{A, X\}$. The adjacency matrix, $A$, integrates $A_{original}$ and $m_g$ as outlined in Equation~\ref{eqn: adj_matrix}. The matrix $A_{original}$, defined as $A_{original\_ij}=\{0, if i=j; 1, otherwise\}$, is fixed and not subject to training, structured in the dimension of $\mathbb{R}^{64\times64}$. Meanwhile, the adjacency matrix mask $m_g \in \mathbb{R}^{64\times64}$ is designated as trainable.

\begin{equation}
\label{eqn: adj_matrix}
A = A_{original}\odot m_g
\end{equation}

\begin{algorithm}[!ht]
\caption{Finding Optimal EEG\_GCN Weights ($\Theta$) and Adjacency Matrix ($m_g$) at Different Density Levels}
\label{alg: EEG_GLT algorithm}

\noindent \textbf{Input:} \parbox[t]{\dimexpr\linewidth-\algorithmicindent}{Graph $G=\{A, X\}$, GNN $f(G,\Theta)$, GNN initialisation \newline $\Theta_{0}$, $A_{original\_ij}= \{0, if \ \ i=j; 1, otherwise\}$, \newline initial Adjacency Matrix Mask $m_{g}^{0}=A_{original}$, \newline learning rate $\eta=0.01$, pruning rate $p_{g}=10 \%$, \newline pre-defined lowest Graph Density Level $s_{g}=13.39\%$. \strut}

\noindent \textbf{Output:} \parbox[t]{\dimexpr\linewidth-\algorithmicindent}{Optimal EEG\_GCN weights ($\Theta^{s}$) with optimal \newline adjacency matrix mask ($m_{g}^{s}$) at different graph \newline density levels.\strut}

\begin{algorithmic}[1]
\While {$\frac{||m_g^s||_0}{||A_{original}||_0}\geq s_g$}
\For{for iteration $i=0, 1, 2, ..., N_{ep}$}
\State \parbox[t]{\dimexpr\linewidth-\algorithmicindent}{Forward $f(.,\Theta^{s}_{i})$ with $G_{s}=\{m_{g}^{s,i}\odot A_{original}, X\}$ \newline to compute Cross-Entropy Loss, $L$ \strut}

\State Backpropagate and update, $\Theta^{s}_{i}$ and $m_{g}^{s,i}$ using Adam Optimiser
\EndFor

\State \parbox[t]{\dimexpr\linewidth-\algorithmicindent} {Record $m_{g}^{s,i}$ and $\Theta_{i}^{s}$ with the highest accuracy in validation set during the $N_{ep}$ iteration \strut}
\State \parbox[t]{\dimexpr\linewidth-\algorithmicindent}{Set $p_{g}=10 \%$ of the lowest absolute magnitude values in $m_{g}^{s}$ to 0 and the others to 1, then obtain a new $m_{g}^{s+1,0}$ \strut}

\EndWhile

\end{algorithmic}
\end{algorithm}

\begin{table*}[!ht]
\caption{Details of EEG\_GLT-Net Model}
\label{table: model eeg_glt-net}
\centering
\setlength{\tabcolsep}{3pt}
\begin{tabular}{m{1.5cm}<{\centering} m{3.0cm} <{\centering} m{1.5cm}<{\centering} m{1.5cm}<{\centering} m{2.5cm}<{\centering} m{1.5cm}<{\centering} }
\hline
\hline
Layer & Type & Input Size & Polynomial Order & Weights & Output \\ \hline
Input & Input & $64\times 1$ & - & - & - \\ 
\hline
\multicolumn{6}{c}{GCN Blocks} \\
\hline
GC1 & Graph Convolution & $64\times 1$ & $5$ & $1\times 16 \times 5$ & $64\times 16$ \\
BNC1 & Batch Normalisation & $64\times 16$ & - & $16$ & $64\times 16$ \\
GC2 & Graph Convolution & $64\times 16$ & $5$ & $16\times 32 \times 5$ & $64\times 32$ \\
BNC2 & Batch Normalisation & $64\times 32$ & - & $32$ & $64\times 32$ \\
GC3 & Graph Convolution & $64\times 32$ & $5$ & $32\times 64 \times 5$ & $64\times 64$ \\
BNC3 & Batch Normalisation & $64\times 64$ & - & $64$ & $64\times 64$ \\
GC4 & Graph Convolution & $64\times 64$ & $5$ & $64\times 128 \times 5$ & $64\times 128$ \\
BNC4 & Batch Normalisation & $64\times 128$ & - & $128$ & $64\times 128$ \\
GC5 & Graph Convolution & $64\times 128$ & $5$ & $128\times 256 \times 5$ & $64\times 256$ \\
BNC5 & Batch Normalisation & $64\times 256$ & - & $256$ & $64\times 256$ \\
GC6 & Graph Convolution & $64\times 256$ & $5$ & $256\times 512 \times 5$ & $64\times 512$ \\
BNC6 & Batch Normalisation & $64\times 512$ & - & $512$ & $64\times 512$ \\
\hline
\multicolumn{6}{c}{Global Mean Pooling Block} \\
\hline
P & Global Mean Pool & $64\times 512$ & - & - & $512$ \\
\hline
\multicolumn{6}{c}{Fully Connected Blocks} \\
\hline
FC1 & Fully Connected & $512$ & - & $512 \times 1024$ & $1024$ \\
BNFC1 & Batch Normalisation & $1024$ & - & $1024$ & $1024$ \\
FC2 & Fully Connected & $1024$ & - & $1024 \times 2048$ & $2048$  \\
BNFC2 & Batch Normalisation & $2048$ & - & $2048$ & $2048$ \\
FC3 & Fully Connected & $2048 \times 4$ & - & $2048 \times 4$ & $4$ \\
S & Softmax Classification & $4$ & - & - & $4$ \\
\hline
\hline

\end{tabular}
\end{table*}

\begin{table}[!ht]
\caption{Hyperparameter Configuration for Training the EEG\_GLT-Net}
\label{table: eeg_glt_net hyperparamter}
\centering
\setlength{\tabcolsep}{3pt}
\begin{tabular}{m{3.5cm}<{\centering} m{1.5cm} <{\centering}}
\hline
\hline
Hyperparamter & Value \\
\hline
Training Epochs $(N_{ep})$ & 1000 \\
Batch Size $(B)$ & 1024 \\
Dropout Rate & 0.5 \\
Optimiser & Adam \\
Initial Learning Rate $(\eta)$ & 0.01 \\
\hline
\hline

\end{tabular}
\end{table}

EEG MI signals from individual subjects, recorded between $t=1s$ and $t=3s$, are trained using Algorithm~\ref{alg: EEG_GLT algorithm}. The detailed structure of the EEG\_GLT-Net is depicted in Figure~\ref{fig: eeg_glt_net_overview} and Table~\ref{table: model eeg_glt-net}, with the specific hyperparameter configurations for the training outlined in Table~\ref{table: eeg_glt_net hyperparamter}. The optimally trained GNN weights ($\Theta$) and the trained adjacency matrix mask ($m_g$) are recorded across various adjacency matrix density levels, ranging from 100\% to 13.39\%.

\subsubsection{EEG MI Time Points GNN Features}
\label{sec: EEG time point features}
From the pre-trained GNN weights and optimal adjacency matrices across varying $m_g$ densities ranging from 100\% to 13.39\%, the set corresponding to a density of 13.39\% was chosen. This density was used to extract graph features from EEG MI signals at specific time points, due to its computation efficiency and superior accuracy compared to the 100\% set. GNN features were then extracted for all EEG MI time points, spanning from t = 0s to t = 4s for all 84 trials of each subject, was conducted. The GNN feature corresponding to each time point had a dimensionality of $\mathbb{R}^{512}$.

\subsection{Problem Redefinition}
\begin{figure*}
    \centering
    \includegraphics[width=0.95\textwidth]{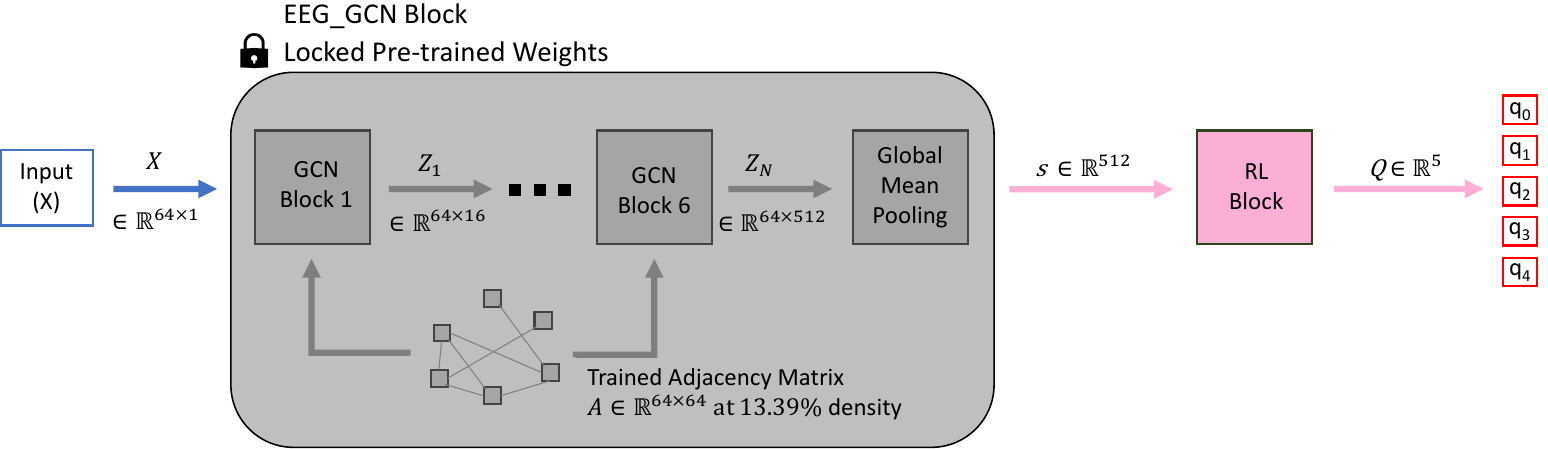}
    \caption{Overview of the EEG\_RL-Net model: Incorporation of the pre-trained EEG\_GCN Block at a 13.39\% $m_g$ density from the EEG\_GLT-Net, coupled with an RL Block}
    \label{fig: eeg_glt_rl_overview}
\end{figure*}

The EEG\_GLT-Net underwent training for the classification of EEG MI time-point signals. Integration the GNN and an optimally trained adjacency matrix significantly enhanced the classification accuracy compared to traditional PCC adjacency matrix method. Nonetheless, ambiguities in signal clarity between different classes at certain time points could adversely affect the model accuracy. Leveraging the high efficacy of the EEG\_GLT-Net model, the pre-trained weights from the GNN and adjacency matrix components were integrated with an RL (Reinforcement Learning) block, resulting in the formation of the EEG\_RL-Net, as depicted in Figure~\ref{fig: eeg_glt_rl_overview}.

\begin{figure}
    \centering
    \includegraphics[width=0.35\textwidth]{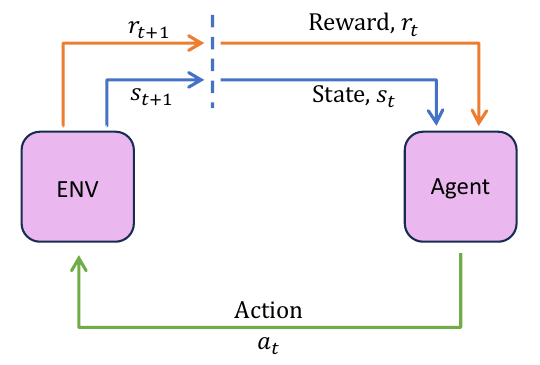}
    \caption{Agent interaction with EEG\_RL Environment}
    \label{fig: eeg_rl env}
\end{figure}

\begin{algorithm}[!ht]
\caption{EEG\_RL Environment}
\label{alg: EEG_RL Environment}
\begin{algorithmic}[1]
\Function{STEP}{$s_t, a_t, y_t, s'_t$}
    \If{$a_t=0$}
        \State $r_t=-0.1$
        \State Return($s'_t, r_t$)
    \Else
        \State $r_t = 
        \begin{cases} 
        r_{right}, \text{ eg. }+10 & \text{if } a_t = y_t \\
        r_{wrong}, \text{ eg. }-10 & \text{if } a_t \neq y_t
        \end{cases}$
        \State Return ($s'_t = Terminal, r_t$)
    \EndIf
\EndFunction
\end{algorithmic}
\end{algorithm}

A reinforcement learning approach is used to train an RL agent for classifying EEG MI time-point signals. Beyond the four initial classes, the RL agent has the capability to defer classification of a current time point if it determines that it is not ready. In each state $s_t$, the RL agent can perform one of five discrete actions $a_t\in\{0, 1, 2, 3, 4\}$ within the EEG\_RL environment, guided by the GNN features extracted from $s_t$. The actions $a_t$ are described as follows:
\begin{itemize}
\item $a_{t}=0:$ Skip the current state $s_t$
\item $a_{t}=1:$ Classify the signal as Class 1
\item $a_{t}=2:$ Classify the signal as Class 2
\item $a_{t}=3:$ Classify the signal as Class 3
\item $a_{t}=4:$ Classify the signal as Class 4
\end{itemize}

Following action $a_t$, the RL agent is rewarded with $r_t$ and transitions to the next state $s_t'$, as illustrated in Figure~\ref{fig: eeg_rl env}. Choosing $a_t=0$ indicates the agent's hesitance to classify due to uncertainty, leading to a decision to skip the current state with a minimal penalty until it is deemed ready to classify or the episode ends. Upon selecting an action $a_t>0$, $s_t'$ is marked as $Terminal$, which concludes the episode and the agent receives $r_t$, a positive reward ($r_{right}$) for correct classification or a negative reward ($r_{wrong}$) for incorrect classification. The dynamics of the EEG\_RL environment are elaborated in Algorithm~\ref{alg: EEG_RL Environment}. The ultimate goal is for the RL agent to accurately classify EEG MI signals within the designated horizon $H=20$ (120 milliseconds) as swiftly as possible.

\subsection{Data Preprocessing and Data Splitting}
\begin{figure}
    \centering
    \includegraphics[width=0.49\textwidth]{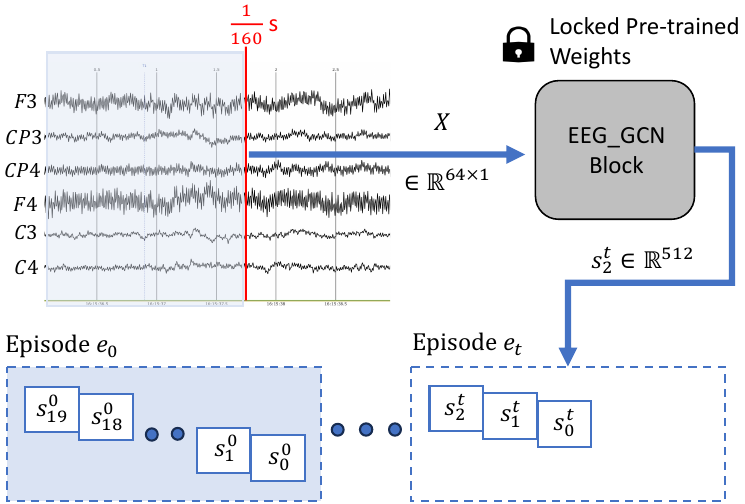}
    \caption{Conversion of EEG MI time points into states using the pre-trained EEG\_GCN Block, grouped into episodes comprising 20 states each}
    \label{fig: episode_form}
\end{figure}

The EEG\_RL-Net training also utilised the PhysioNet dataset, consistent with the approach for EEG\_GLT-Net training. For this training, the entire duration of the EEG MI signals was included, spanning four seconds at a sampling rate of 160Hz, was included. As outlined in Section~\ref{sec: EEG time point features}, GNN features of EEG MI time points were used, which were extracted by leveraging pre-trained weights at a 13.39\%. The GNN features of each time point were considered as states, $s\in\mathbb{R}^{512}$. For all 82 trials, from $t=0s$ to $t=4s$, groups of consecutive $H=20$ states were organised into episodes without time point overlap between subsequent episodes, creating an episode set, $E\in\{e_0, e_1, ..., e_n\}$, as illustrated in Figure~\ref{fig: episode_form}. Episodes were randomised using a specific seed, with 80\% of the episodes in $E$ allocated for the training set ($E_{train}$), 10\% for the validation set ($E_{val}$), and the remaining 10\% for the test set ($E_{test}$).

\subsection{Dueling Deep Q-Learning}
The Deep Q Learning Network (DQN) method, a value-based RL approach, was employed in this study to learn an optimal policy to enable more accurate classification of EEG MI signals. A state-action value, $Q(s,a)$, represents the expected discounted reward when the agent is in state $s$, and takes action $a$ according to policy $\pi$. With the optimal policy ($\pi^{*}$), the agent aims to achieve the maximum expected discounted reward $Q^{*}(s,a)$, fulfilling the Bellman equation: 

\begin{equation}
\label{eqn: bellman}
Q^{*}(s,a)=\mathbb{E}_{\pi^*}[r+\gamma 
 \text{max}_{a'}Q^{*}(s',a')|s,a]
\end{equation}  
\noindent here $r$ is the immediate reward, and $\gamma$ is the discount factor. The state-action value, $\hat{Q}(s,a)$, for state $s$ and action $a$ can be approximated using a deep neural network parameterised by $\theta$. The loss function is defined as:

\begin{equation}
\label{eqn: rl-loss}
Loss(\theta)= (\hat{y}^{DQN} - \hat{Q}(s,a;\theta))^2
\end{equation}

\noindent where $\hat{y}^{DQN}$ is the target value, calculated as follows:

\begin{equation}
\label{eqn: y_dqn}
\hat{y}^{DQN} =
\begin{cases}
    r_t, & \text{if } s'_t \text{ is Terminal} \\
    r_t + \gamma \max_{a'_t} \hat{Q}(s'_t, a'_t; \theta_{\text{target}}), & \text{otherwise}
\end{cases}
\end{equation}

The $\theta_{target}$ denotes the parameters of the target network, which are kept constant. The approximation $\hat{Q}(s,a;\theta)$ shares the architecture with the target network. Our study utilises Dueling DQN, a variant of DQN that enhances training stability and efficiency by separating the estimation of $\hat{Q}(s,a;\theta)$ into state values $V(s)$ and action advantages $A(s,a)$, as follows:

\begin{equation}
\label{eqn: duel-dqn-1}
\hat{Q}(s,a;\theta)=\hat{V}(s;\alpha)+\hat{A}(s,a;\beta)
\end{equation}

The network separately estimates the state values and action advantages, which then converge into a single output. The parameters $\theta$ represent the overall network parameters, with $\alpha$ and $\beta$ specifically used for estimating state values and action advantages, respectively. To enhance stability, the equation subtracts the average advantage values from $\hat{Q}(s,a;\theta)$:

\begin{equation}
\label{eqn: duel-dqn-2}
\hat{Q}(s,a;\theta)=\hat{V}(s;\alpha)+[\hat{A}(s,a;\beta)-\frac{1}{|A|}\sum_{a}\hat{A}(s,a;\beta)]
\end{equation}

\begin{figure}
    \centering
    \includegraphics[width=0.495\textwidth]{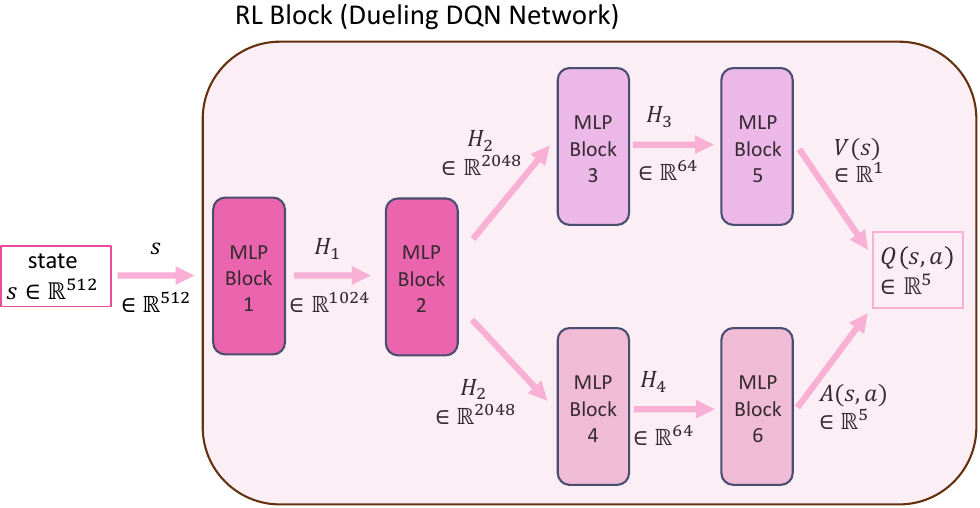}
    \caption{EEG\_RL-Net's RL Block: Featuring the Dueling Deep Q Network (DQN), this component predicts the q-values linked to various actions}
    \label{fig: rl_block}
\end{figure}

\subsection{EEG\_RL Algorithm}
To generate training data for the RL Block, all possible actions $a_t=\{0, 1, 2, 3, 4\}$ are executed at each state $s_t$ within an episode $e_i$ in $E_{train}$, interacting with the EEG\_RL environment to determine the reward $r_t$ and the subsequent state $s'_t$. Each transition records a tuple ($s, a, r, s'$). This study employs the Dueling DQN method for the RL block, as illustrated in Figure~\ref{fig: rl_block}. The Dueling DQN agent undergoes training according to the procedure outlined in Algorithm~\ref{alg: DQN_Train}, utilising the Adam optimiser until convergence is achieved. The configuration of the entire EEG\_RL-Net model is outlined in Table~\ref{table: eeg_rl_net model details}. The parameters of the fixed target network, $\theta_{target}$, for the Dueling DQN network, are refreshed after every 50 batch updates of $\theta$.

\begin{algorithm}[!ht]
\caption{Training EEG\_RL-Net's Dueling DQN Agent}
\label{alg: DQN_Train}
\begin{algorithmic}[1]
\State Initialise randomly Dueling DQN network parameter ($\theta$) and target network parameter ($\theta_{target}$).
\State Set of train episodes $E_{TRAIN}\in \{e_0, e_1, ..., e_N\}$ where each $e_i$ has set of states, $S=\{s_0, s_1, ..., s_{H-1}\}$. Each state, $s_t\in \mathbb{R}^{512}$.
\State At each state $s_t$, simulate one step with all possible actions from action set, $A\in \{0, 1, 2, 3, 4\}$ to observe next state, $s'_t$ and reward, $r_t$. Record all the ($s_{t}, a_{t}, r_{t}, s'_{t}$) tuples to the Buffer $B$.
\State Shuffle the state transitions in the $B$ using random seed, and group into mini-batches in size of 64 transitions.

\For {$epoch=0$ to $EPOCHS$}
    \State Compute $\hat{y}^{DQN}$ for each mini-batch:
    \State $\hat{y}^{DQN} =
        \begin{cases}
        r_t, & \hspace{-10mm}\text{if } s'_t \text{ is Terminal} \\
        r_t + \gamma \max_{a'_t} \hat{Q}(s'_t, a'_t; \theta_{\text{target}}) & \text{otherwise}
        \end{cases}$
        
    \State $Loss(\theta)=(\hat{y}^{DQN} - \hat{Q}(s_t, a_t; \theta))^2$
    \State Backpropagate to update $\theta$ using $Adam$ optimiser
    \State Update $\theta_{target}= \theta$ at every 50 updates of $\theta$        
\EndFor

\end{algorithmic}
\end{algorithm}

Performance evaluation of the RL agent on $E_{val}$ and $E_{test}$ is described in Algorithm~\ref{alg: DQN_Test Eval}. At every time step, the agent selects an action based on the q-values predicted by the EEG\_RL-Net. In this study, a correct classification by the agent yields a reward $r_{right}$, while an incorrect classification results in $r_{wrong}$. The agent's objective in each episode $e_i$ is to maximise the cumulative reward $r_{sum}$ within the predefined horizon $H=20$. This requires the agent to make classifications as quickly as possible, since it incurs a penalty of $r=-0.1$ for each skipped step. However, at time $t=H-1$, skipping is no longer an option, and the agent must make a classification action.

\begin{algorithm}[!ht]
\caption{Evaluation of DQN Agent for a Validation or Test Episode}
\label{alg: DQN_Test Eval}
\begin{algorithmic}[1]
\State Episode, $e_i$ has horizon of $H=20$
\State At $e_i$, the set of states $S=\{s_0, s_1, ..., s_{H-1}\}$, where each $s_t\in\mathbb{R}^{512}$
\State At $e_i$, the set of labels $Y=\{y_0, y_1, ..., y_{H-1}\}$, where each $y_t\in\{1, 2, 3, 4\}$
\State Action $a'\in \{0, 1, 2, 3, 4\}$, and $a''\in \{1, 2, 3, 4\}$
\State Initialise $t=0$, $r_{sum}=0$
\While {$t<H$}
    \State $a_t = 
        \begin{cases} 
        argmax_{a'}\hat{q}_{DQN}(s_t, a'), & \text{if } t<H-1 \\
        argmax_{a''}\hat{q}_{DQN}(s_t, a''), & \text{otherwise }
        \end{cases}$

    \State $s'_t, r_t = STEP (s_t, a_t, y_t, s'_t$)
    \State $r_{sum}\gets r_{sum} + r_t$
    \If {$r_t = Terminate$}
        \State Terminate the Episode, $e_i$
    \Else  
        \State $t \gets t + 1$   
    \EndIf
\EndWhile

\end{algorithmic}
\end{algorithm}

\subsection{Model Setting and Evaluation Metrics}
The structure of EEG\_RL-Net is defined by two principal components: the spectral EEG\_GCN block, which extracts graph features from EEG MI time point signals using pre-trained weights, and the RL block, embodied by the Dueling DQN network. The specifics of the EEG\_RL-Net's design are provided in Table~\ref{table: eeg_rl_net model details}. The RL block comprises six MLP (Multi-Layer Perceptron) layers, or Fully Connected Layers, each followed by a Rectified Linear Unit (ReLU) layer, as described in Equation~\ref{eqn: relu}. Information on the training hyperparameters is presented in Table~\ref{table: hyperparamter-eegrl}. The performance of the different methods was evaluated using both accuracy and F1 score metrics.

\begin{table}[!ht]
\caption{Details of EEG\_RL-Net Model}
\label{table: eeg_rl_net model details}
\centering
\setlength{\tabcolsep}{3pt}
\begin{tabular}{m{1.5cm}<{\centering} m{2.2cm} <{\centering} m{1.5cm}<{\centering} m{1.5cm}<{\centering} m{1.0cm}<{\centering}}
\hline
\hline
Layer & Type & Input Size &  Weights & Output \\ 
\hline
Input & Input & $64\times 1$ & - & - \\ 
\hline
\multicolumn{5}{c}{EEG\_GCN Block} \\
\hline
EEG\_GCN & Graph Convolution and Global Pooling & $64\times1$ & - & $512$ \\
\hline

\multicolumn{5}{c}{RL Block (Dueling DQN Network)} \\
\hline
MLP1 & Fully Connected & $512$  & $512 \times 1024$ & $1024$ \\
MLP2 & Fully Connected & $1024$ & $1024 \times 2048$ & $2048$ \\
MLP3 & Fully Connected & $2048$ & $2048 \times 64$ & $64$ \\
MLP4 & Fully Connected & $64$  & $64 \times 1$ & $1$ \\
MLP5 & Fully Connected & $2048$ & $2048 \times 64$ & $64$ \\
MLP6 & Fully Connected & $64$ & $64 \times 5$ & $5$ \\
Q & Dueling DQN & $64\times 1$ \& $64\times 5$  & - & $5$ \\

\hline
\hline
\end{tabular}
\end{table}

\begin{table}[!ht]
\caption{Hyperparameter Configuration for Training the EEG\_RL-Net}
\label{table: hyperparamter-eegrl}
\centering
\setlength{\tabcolsep}{3pt}
\begin{tabular}{m{4.0cm}<{\centering} m{1.5cm} <{\centering}}
\hline
\hline
Hyperparamter & Value \\
\hline
Reward Right $(r_{right})$ & $+10$ \\
Reward Wrong $(r_{wrong})$ & $-10$ \\
Reward Skip $(r_{skip})$ & $-0.1$ \\
Discount Factor $(\gamma)$ & 0.99 \\
Training Epoch $(EPOCHS)$ & 150 \\
Batch Size & 63 \\
Target Network Update Frequency & 50 \\
Initial Learning Rate $(\eta)$ & 0.0001 \\
L2 Regularisation Rate $(\lambda)$ & 0.001 \\
Optimiser & Adam \\

\hline
\hline

\end{tabular}
\end{table}

\begin{equation}
\label{eqn: relu}
ReLU(x)=max(0, x)
\end{equation}

\begin{equation}
\label{eqn: accuracy}
Accuracy=\frac{TP + TN}{TP + FP + TN + FN}
\end{equation}

\begin{equation}
\label{eqn: sensitivity}
Sensitivity=\frac{TP}{TP + FN}
\end{equation}

\begin{equation}
\label{eqn: precision}
Precision=\frac{TP}{TP + FP}
\end{equation}

\begin{equation}
\label{eqn: f1 score}
F1\ Score=\frac{2\times Precision\times Sensitivity}{Precision + Sensitivity}
\end{equation}

\section{Results and Discussion}
\subsection{EEG\_RL-Net vs EEG\_GLT-Net}

Table~\ref{table: eeg_glt vs eeg_rl} shows comparative analysis of mean accuracy between the EEG\_GLT-Net and the EEG\_RL-Net. The EEG\_GLT-Net incorporates two adjacency matrix types: the Pearson Coefficient Correlation (PCC) and the $m_{g\_GLT}$. The latter is identified as the most optimal adjacency matrix, after searching through 100\% to 13.39\% of adjacency matrix density using the EEG\_GLT algorithm. According to paper \cite{aung2024eeggltnet}, employing the $m_{g\_GLT}$ adjacency matrix yields an accuracy improvement ranging between 0.51\% and 22.04\% over the PCC adjacency matrix, with significant enhancements noted for subjects $S_1$ and $S_{12}$, at 22.04\% and 21.62\% respectively. Despite seeing notable improvements in accuracy and F1 score with the $m_{g\_GLT}$ matrix, certain subjects, specifically $S_5$, $S_6$, $S_7$, $S_{13}$, $S_{15}$, and $S_{19}$, exhibited classification accuracies below 70\%.

\begin{table}[ht]
\caption{Accuracy Assessment: EEG\_RL-Net versus EEG\_GLT-Net}
\label{table: eeg_glt vs eeg_rl}
\centering
\setlength{\tabcolsep}{3pt}
\begin{tabular}{m{1.0cm}<{\centering} m{2.2cm}<{\centering} m{2.2cm}<{\centering} m{2.5cm}<{\centering}}
\hline
\hline

\multirow{2}{*}{Subj} & \multicolumn{3}{c}{Accuracy (Mean$\pm$Std)}\\ 
\cline{2-4}
& EEG\_GLT-Net\newline(PCC Adj) & EEG\_GLT-Net \newline($m_{g\_GLT}$ Adj) & EEG\_RL-Net* \newline(our method)\\
\hline

$S_{1}$ & \multicolumn{1}{r}{76.47\% $\pm$ 9.94\%} & \multicolumn{1}{r}{98.51\% $\pm$ 0.77\%} & \multicolumn{1}{r}{\textbf{100.00\% $\pm$ 0.00\%}} \\
$S_{2}$ & \multicolumn{1}{r}{69.13\% $\pm$ 7.05\%} & \multicolumn{1}{r}{76.18\% $\pm$ 5.53\%} & \multicolumn{1}{r}{\textbf{97.73\% $\pm$ 0.20\%}} \\
$S_{3}$ & \multicolumn{1}{r}{87.28\% $\pm$ 9.19\%} & \multicolumn{1}{r}{99.17\% $\pm$ 0.32\%} & \multicolumn{1}{r}{\textbf{100.00\% $\pm$ 0.00\%}} \\
$S_{4}$ & \multicolumn{1}{r}{99.13\% $\pm$ 1.01\%} & \multicolumn{1}{r}{99.97\% $\pm$ 0.06\%} & \multicolumn{1}{r}{\textbf{100.00\% $\pm$ 0.00\%}} \\
$S_{5}$ & \multicolumn{1}{r}{43.19\% $\pm$ 3.03\%} & \multicolumn{1}{r}{50.95\% $\pm$ 3.80\%} & \multicolumn{1}{r}{\textbf{87.72\% $\pm$ 0.70\%}} \\
$S_{6}$ & \multicolumn{1}{r}{58.23\% $\pm$ 5.19\%} & \multicolumn{1}{r}{69.60\% $\pm$ 5.67\%} & \multicolumn{1}{r}{\textbf{90.89\% $\pm$ 1.50\%}} \\
$S_{7}$ & \multicolumn{1}{r}{50.98\% $\pm$ 3.80\%} & \multicolumn{1}{r}{59.45\% $\pm$ 3.00\%} & \multicolumn{1}{r}{\textbf{89.24\% $\pm$ 2.10\%}} \\
$S_{8}$ & \multicolumn{1}{r}{95.06\% $\pm$ 5.96\%} & \multicolumn{1}{r}{99.95\% $\pm$ 0.07\%} & \multicolumn{1}{r}{\textbf{100.00\% $\pm$ 0.00\%}} \\
$S_{9}$ & \multicolumn{1}{r}{97.64\% $\pm$ 3.33\%} & \multicolumn{1}{r}{99.95\% $\pm$ 0.08\%} & \multicolumn{1}{r}{\textbf{100.00\% $\pm$ 0.00\%}} \\
$S_{10}$ & \multicolumn{1}{r}{99.24\% $\pm$ 0.19\%} & \multicolumn{1}{r}{99.99\% $\pm$ 0.01\%} & \multicolumn{1}{r}{\textbf{100.00\% $\pm$ 0.00\%}} \\
$S_{11}$ & \multicolumn{1}{r}{99.48\% $\pm$ 0.70\%} & \multicolumn{1}{r}{99.99\% $\pm$ 0.01\%} & \multicolumn{1}{r}{\textbf{100.00\% $\pm$ 0.00\%}} \\
$S_{12}$ & \multicolumn{1}{r}{78.07\% $\pm$ 8.95\%} & \multicolumn{1}{r}{99.69\% $\pm$ 0.32\%} & \multicolumn{1}{r}{\textbf{100.00\% $\pm$ 0.00\%}} \\
$S_{13}$ & \multicolumn{1}{r}{41.35\% $\pm$ 1.23\%} & \multicolumn{1}{r}{44.50\% $\pm$ 2.23\%} & \multicolumn{1}{r}{\textbf{89.45\% $\pm$ 0.90\%}} \\
$S_{14}$ & \multicolumn{1}{r}{55.97\% $\pm$ 6.47\%} & \multicolumn{1}{r}{72.39\% $\pm$ 6.43\%} & \multicolumn{1}{r}{\textbf{91.59\% $\pm$ 2.10\%}} \\
$S_{15}$ & \multicolumn{1}{r}{52.11\% $\pm$ 3.96\%} & \multicolumn{1}{r}{67.55\% $\pm$ 9.26\%} & \multicolumn{1}{r}{\textbf{80.83\% $\pm$ 1.50\%}} \\
$S_{16}$ & \multicolumn{1}{r}{96.75\% $\pm$ 5.00\%} & \multicolumn{1}{r}{99.98\% $\pm$ 0.03\%} & \multicolumn{1}{r}{\textbf{100.00\% $\pm$ 0.00\%}} \\
$S_{17}$ & \multicolumn{1}{r}{98.83\% $\pm$ 2.33\%} & \multicolumn{1}{r}{99.98\% $\pm$ 0.03\%} & \multicolumn{1}{r}{\textbf{100.00\% $\pm$ 0.00\%}} \\
$S_{18}$ & \multicolumn{1}{r}{86.19\% $\pm$ 9.95\%} & \multicolumn{1}{r}{99.92\% $\pm$ 0.12\%} & \multicolumn{1}{r}{\textbf{100.00\% $\pm$ 0.00\%}} \\
$S_{19}$ & \multicolumn{1}{r}{38.38\% $\pm$ 2.27\%} & \multicolumn{1}{r}{41.41\% $\pm$ 1.44\%} & \multicolumn{1}{r}{\textbf{79.65\% $\pm$ 1.40\%}} \\
$S_{20}$ & \multicolumn{1}{r}{98.44\% $\pm$ 0.68\%} & \multicolumn{1}{r}{99.94\% $\pm$ 0.11\%} & \multicolumn{1}{r}{\textbf{100.00\% $\pm$ 0.00\%}} \\
\hline
Overall & \multicolumn{1}{r}{76.10\% $\pm$ 22.71\%} & \multicolumn{1}{r}{83.95\% $\pm$ 21.43\%} & \multicolumn{1}{r}{\textbf{95.36\% $\pm$ 6.83\%}} \\

\hline
\hline
\multicolumn{4}{l}{* $r_{right}=+10$, $r_{wrong}=-10$, $r_{skip}=-0.1$, $H=20$}\\

\end{tabular}
\end{table}

Using baseline parameters ($r_{right}=+10$, $r_{wrong}=-10$, $r_{skip}=-0.1$ and $H=20$), the EEG\_RL-Net framework advances the accuracy beyond the current state-of-the-art EEG\_GLT-Net employing the $m_{g\_GLT}$ adjacency matrix, with improvements spanning 0.01\% to 44.95\%. A total of 12 out of 20 subjects, namely $S_1$, $S_3$, $S_4$, $S_8$, $S_9$, $S_{10}$, $S_{11}$, $S_{12}$, $S_{16}$, $S_{17}$, $S_{18}$, and $S_{20}$, achieved perfect classification. The EEG\_RL-Net also significantly elevated the accuracies for $S_{13}$ and $S_{19}$ to 89.45\% and 79l.65\%, respectively. Even for subjects $S_{13}$ and $S_{19}$, who initially demonstrated low accuracies, modest improvement in accuracy at 44.50\% and 41.41\%, respectively was achieved using the EEG\_GLT-Net with the $m_{g\_GLT}$ matrix.

The EEG\_GLT-Net with the $m_{g\_GLT}$ matrix boosted accuracy across the 20 subjects, increasing the average accuracy by 7.85\% (from 76.10\% to 83.95\%). Given the inherent noise in EEG MI time-point signals and the challenge of classifying signals representing $\frac{1}{160}s$, the EEG\_GLT-Net showed a decline in performance accuracy due to its attempt to classify all time points. Comparatively, the EEG\_RL-Net achieved remarkable increase in average accuracy across the 20 subjects to 95.35\%. This substantial improvement is the result of the RL agent's capacity to discern the appropriateness of the current signal for classification. The agent has been optimised to classify signals as swiftly as possible within a 20 time-point window, averaging a classification time of 2.91 time points in the EEG\_RL-Net setup.

\subsection{Study of Changing $r_{right}$ Values}

Table~\ref{table: r_right vary} demonstrates the effect of varying the $r_{right}$ value ($+5, +10, +15$), on average accuracy while keeping $r_{wrong}=-10$ constant. The results show average accuracies of 95.57\%, 95.36\%, and 94.94\% for $r_{right}=+5, +10,$ and $+20$, respectively. Notably, the accuracy tends to improve when $r_{right}$ is less than $r_{wrong}$, but declines when $r_{right}$ exceeds $r_{wrong}$, although the level of variance is minimal at just 0.63\%.

\begin{table}[ht]
\caption{Impact of Varying $r_{right}$ Values on Accuracy and Classification Time}
\label{table: r_right vary}
\centering
\setlength{\tabcolsep}{2pt}
\begin{tabular}{m{0.6cm}<{\centering} m{2.0cm}<{\centering} m{2.0cm}<{\centering} m{2.0cm}<{\centering} }

\hline
\hline

\multirow{3}{*}{Subj} & \multicolumn{3}{c}{Mean Accuracy (Mean Classification Time)}\\
\cline{2-4} 
 & $r_{r}=+5$\newline$r_{w}=-10$ & $r_{r}=+10$\newline$r_{w}=-10$ & $r_{r}=+20$\newline$r_{w}=-10$ \\
 
\hline
$S_1$ &  \multicolumn{1}{r}{99.93\% (1.70)}  & \multicolumn{1}{r}{100.00\% (1.80)}  & \multicolumn{1}{r}{100.00\% (1.50)} \\
$S_2$ &  \multicolumn{1}{r}{97.86\% (1.51)} & \multicolumn{1}{r}{97.73\% (1.87)} & \multicolumn{1}{r}{97.86\% (1.65)} \\
$S_3$ &  \multicolumn{1}{r}{100.00\% (2.10)} & \multicolumn{1}{r}{100.00\% (2.20)} & \multicolumn{1}{r}{100.00\% (1.90)} \\
$S_4$ &  \multicolumn{1}{r}{100.00\% (2.80)} & \multicolumn{1}{r}{100.00\% (2.50)} & \multicolumn{1}{r}{100.00\% (1.80)} \\
$S_5$ &  \multicolumn{1}{r}{87.65\% (5.71)} & \multicolumn{1}{r}{87.72\% (4.55)} & \multicolumn{1}{r}{86.14\% (3.45)} \\
$S_6$ &  \multicolumn{1}{r}{91.10\% (3.37)} & \multicolumn{1}{r}{90.89\% (2.80)} & \multicolumn{1}{r}{89.72\% (2.07)} \\
$S_7$ &  \multicolumn{1}{r}{91.24\% (3.64)} & \multicolumn{1}{r}{89.24\% (3.23)} & \multicolumn{1}{r}{87.80\% (3.66)} \\
$S_8$ &  \multicolumn{1}{r}{100.00\% (2.90)} & \multicolumn{1}{r}{100.00\% (2.80)} & \multicolumn{1}{r}{100.00\% (2.10)} \\
$S_9$ &  \multicolumn{1}{r}{100.00\% (3.70)} & \multicolumn{1}{r}{100.00\% (3.00)} & \multicolumn{1}{r}{100.00\% (2.20)} \\
$S_{10}$ &  \multicolumn{1}{r}{99.93\% (2.40)} & \multicolumn{1}{r}{100.00\% (2.20)} & \multicolumn{1}{r}{100.00\% (1.80)} \\
$S_{11}$ &  \multicolumn{1}{r}{100.00\% (2.00)} & \multicolumn{1}{r}{100.00\% (1.50)} & \multicolumn{1}{r}{100.00\% (2.30)} \\
$S_{12}$ &  \multicolumn{1}{r}{100.00\% (2.80)} & \multicolumn{1}{r}{100.00\% (2.40)} & \multicolumn{1}{r}{100.00\% (2.60)} \\
$S_{13}$ &  \multicolumn{1}{r}{89.59\% (4.90)} & \multicolumn{1}{r}{89.45\% (3.97)} & \multicolumn{1}{r}{88.28\% (3.58)} \\
$S_{14}$ &  \multicolumn{1}{r}{93.45\% (3.75)} & \multicolumn{1}{r}{91.59\% (3.14)} & \multicolumn{1}{r}{89.86\% (2.81)} \\
$S_{15}$ &  \multicolumn{1}{r}{82.89\% (5.23)} & \multicolumn{1}{r}{80.83\% (4.80)} & \multicolumn{1}{r}{79.45\% (4.51)} \\
$S_{16}$ &  \multicolumn{1}{r}{100.00\% (2.40)} & \multicolumn{1}{r}{100.00\% (2.70)} & \multicolumn{1}{r}{100.00\% (2.20)} \\
$S_{17}$ &  \multicolumn{1}{r}{100.00\% (3.00)} & \multicolumn{1}{r}{100.00\% (2.00)} & \multicolumn{1}{r}{100.00\% (1.90)} \\
$S_{18}$ &  \multicolumn{1}{r}{100.00\% (3.30)} & \multicolumn{1}{r}{100.00\% (2.40)} & \multicolumn{1}{r}{100.00\% (1.50)} \\
$S_{19}$ &  \multicolumn{1}{r}{77.79\% (6.89)} & \multicolumn{1}{r}{79.65\% (5.64)} & \multicolumn{1}{r}{79.59\% (4.80)} \\
$S_{20}$ &  \multicolumn{1}{r}{100.00\% (2.20)} & \multicolumn{1}{r}{100.00\% (2.70)} & \multicolumn{1}{r}{100.00\% (2.10)} \\
\hline
Mean & \multicolumn{1}{r}{95.57\% (3.32)} & \multicolumn{1}{r}{95.36\% (2.91)} & \multicolumn{1}{r}{94.94\% (2.51)} \\
Std & \multicolumn{1}{r}{$\pm$ 6.72\%} & \multicolumn{1}{r}{$\pm$ 6.83\%} & \multicolumn{1}{r}{$\pm$ 7.32\%} \\
\hline
\hline
\multicolumn{4}{l}{$r_{skip}=-0.1$ and $H=20$}\\

\end{tabular}
\end{table}

On an individual basis, $r_{right}=+5$ yielded higher accuracies for most subjects, except for $S_1$, $S_5$, and $S_{19}$, where $r_{right}=+10$ performed marginally better. No subjects showed improved accuracy when $r_{right}$ was greater than $r_{wrong}$. Therefore, for optimal performance, the magnitude of $r_{right}$ should not exceed $r_{wrong}$. It appears that accuracy is enhanced by a higher penalty for incorrect classifications ($r_{wrong}$) rather than a higher reward for correct ones ($r_{right}$) enhances accuracy, likely motivating the agent to avoid misclassifications more stringently.

Regarding the time points required to classify EEG MI signals, the configuration with $r_{right}=+10$ and $r_{wrong}=-10$ averages at 2.91 time points. Increasing $r_{right}$ to $+20$ (while $r_{wrong}$ remains at $-10$) reduces the classification time to 2.51 time points. Conversely, lowering $r_{right}$ to $+5$ increases the average classification time to 3.32 time points, indicating a more cautious approach by the agent, likely due to prioritising accuracy over speed by utilising the option to skip uncertain classifications.

\subsection{Study of Changing $r_{wrong}$ Values}

In this study, we examined the impact of altering the $r_{wrong}$ values while keeping the $r_{right}$ constant at $+10$, as shown in Table~\ref{table: r_wrong vary}. We observed the $r_{wrong}$ values at $-10$, $-20$, $-30$, and $-40$, correlating with an average performance accuracy of 95.35\%, 95.18\%, 95.11\%, and 94.88\%, respectively. This indicates that simply increasing the negative magnitude of $r_{wrong}$ beyond that of $r_{right}$ does not invariably lead to enhanced performance accuracy. Additionally, we found that the time required for signal classification was directly related to the difference in rewards.

\begin{table}[ht]
\caption{Impact of Varying $r_{wrong}$ Values on Accuracy and Classification Time}
\label{table: r_wrong vary}
\centering
\setlength{\tabcolsep}{2pt}
\begin{tabular}{m{0.6cm}<{\centering} m{1.8cm}<{\centering} m{1.8cm}<{\centering} m{1.8cm}<{\centering} m{1.8cm}<{\centering}}
\hline
\hline

\multirow{3}{*}{Subj} & \multicolumn{4}{c}{Mean Accuracy (Mean Classification Time)}\\
\cline{2-5} 
 & $r_{r}=+10$\newline$r_{w}=-10$ & $r_{r}=+10$\newline$r_{w}=-20$ & $r_{r}=+10$\newline$r_{w}=-30$ & $r_{r}=+10$\newline$r_{w}=-40$ \\

\hline
$S_1$ &  \multicolumn{1}{r}{100.00\% (1.80)} & \multicolumn{1}{r}{99.79\% (1.60)} & \multicolumn{1}{r}{100.00\% (2.00)} & \multicolumn{1}{r}{99.93\% (2.00)} \\
$S_2$ &  \multicolumn{1}{r}{97.73\% (1.87)} & \multicolumn{1}{r}{98.21\% (1.97)} & \multicolumn{1}{r}{97.93\% (2.49)} & \multicolumn{1}{r}{97.93\% (2.88)} \\
$S_3$ &  \multicolumn{1}{r}{100.00\% (2.20)} & \multicolumn{1}{r}{100.00\% (1.70)} & \multicolumn{1}{r}{100.00\% (2.70)} & \multicolumn{1}{r}{100.00\% (2.20)} \\
$S_4$ &  \multicolumn{1}{r}{100.00\% (2.50)} & \multicolumn{1}{r}{100.00\% (3.30)} & \multicolumn{1}{r}{100.00\% (4.30)} & \multicolumn{1}{r}{100.00\% (5.70)} \\
$S_5$ &  \multicolumn{1}{r}{87.72\% (4.55)} & \multicolumn{1}{r}{86.90\% (6.26)} & \multicolumn{1}{r}{85.38\% (6.28)} & \multicolumn{1}{r}{85.79\% (7.31)} \\
$S_6$ &  \multicolumn{1}{r}{90.89\% (2.80)} & \multicolumn{1}{r}{90.00\% (2.63)} & \multicolumn{1}{r}{90.90\% (4.94)} & \multicolumn{1}{r}{90.41\% (4.25)} \\
$S_7$ &  \multicolumn{1}{r}{89.24\% (3.23)} & \multicolumn{1}{r}{89.59\% (5.09)} & \multicolumn{1}{r}{89.31\% (5.20)} & \multicolumn{1}{r}{88.90\% (6.70)} \\
$S_8$ &  \multicolumn{1}{r}{100.00\% (2.80)} & \multicolumn{1}{r}{100.00\% (2.60)} & \multicolumn{1}{r}{100.00\% (3.70)} & \multicolumn{1}{r}{100.00\% (4.30)} \\
$S_9$ &  \multicolumn{1}{r}{100.00\% (3.00)} & \multicolumn{1}{r}{100.00\% (3.30)} & \multicolumn{1}{r}{100.00\% (5.00)} & \multicolumn{1}{r}{100.00\% (5.30)} \\
$S_{10}$ &  \multicolumn{1}{r}{100.00\% (2.20)} & \multicolumn{1}{r}{99.93\% (1.90)} & \multicolumn{1}{r}{99.86\% (2.00)} & \multicolumn{1}{r}{99.93\% (2.52)} \\
$S_{11}$ &  \multicolumn{1}{r}{100.00\% (1.50)} & \multicolumn{1}{r}{100.00\% (2.20)} & \multicolumn{1}{r}{100.00\% (3.00)} & \multicolumn{1}{r}{100.00\% (3.30)} \\
$S_{12}$ &  \multicolumn{1}{r}{100.00\% (2.40)} & \multicolumn{1}{r}{100.00\% (2.70)} & \multicolumn{1}{r}{100.00\% (4.10)} & \multicolumn{1}{r}{100.00\% (5.10)} \\
$S_{13}$ &  \multicolumn{1}{r}{89.45\% (3.97)} & \multicolumn{1}{r}{89.52\% (5.33)} & \multicolumn{1}{r}{89.52\% (6.24)} & \multicolumn{1}{r}{87.38\% (6.07)} \\
$S_{14}$ &  \multicolumn{1}{r}{91.59\% (3.14)} & \multicolumn{1}{r}{91.31\% (3.00)} & \multicolumn{1}{r}{92.14\% (3.00)} & \multicolumn{1}{r}{90.55\% (4.37)} \\
$S_{15}$ &  \multicolumn{1}{r}{80.83\% (4.80)} & \multicolumn{1}{r}{81.38\% (4.53)} & \multicolumn{1}{r}{80.14\% (3.48)} & \multicolumn{1}{r}{80.69\% (4.13)} \\
$S_{16}$ &  \multicolumn{1}{r}{100.00\% (2.70)} & \multicolumn{1}{r}{100.00\% (3.10)} & \multicolumn{1}{r}{100.00\% (5.00)} & \multicolumn{1}{r}{100.00\% (4.30)} \\
$S_{17}$ &  \multicolumn{1}{r}{100.00\% (2.00)} & \multicolumn{1}{r}{100.00\% (3.70)} & \multicolumn{1}{r}{100.00\% (3.20)} & \multicolumn{1}{r}{100.00\% (4.60)} \\
$S_{18}$ &  \multicolumn{1}{r}{100.00\% (2.40)} & \multicolumn{1}{r}{100.00\% (3.20)} & \multicolumn{1}{r}{100.00\% (2.40)} & \multicolumn{1}{r}{100.00\% (4.30)} \\
$S_{19}$ &  \multicolumn{1}{r}{79.65\% (5.64)} & \multicolumn{1}{r}{76.90\% (7.21)} & \multicolumn{1}{r}{76.97\% (8.22)} & \multicolumn{1}{r}{76.14\% (9.20)} \\
$S_{20}$ &  \multicolumn{1}{r}{100.00\% (2.70)} & \multicolumn{1}{r}{100.00\% (2.80)} & \multicolumn{1}{r}{100.00\% (3.40)} & \multicolumn{1}{r}{100.00\% (4.80)} \\
\hline
Mean & \multicolumn{1}{r}{95.36\% (2.91)} & \multicolumn{1}{r}{95.18\% (3.41)} & \multicolumn{1}{r}{95.11\% (4.03)} & \multicolumn{1}{r}{94.88\% (4.66)} \\
Std & \multicolumn{1}{r}{$\pm$ 6.83\%} & \multicolumn{1}{r}{$\pm$ 7.19\%} & \multicolumn{1}{r}{$\pm$ 7.37\%} & \multicolumn{1}{r}{$\pm$ 7.57\%} \\
\hline
\hline
\multicolumn{5}{l}{$r_{skip}=-0.1$ and $H=20$}\\
\end{tabular}
\end{table}

Despite the reward configuration of $\{r_{right}=+10, r_{wrong} = -10\}$ achieving the highest average performance accuracy among the four settings, it does not universally outperform across all test subjects. Specifically, this configuration was only superior for subjects $S_5$ and $S_6$. Conversely, the configuration of $\{r_{right}=+10, r_{wrong}=-20\}$ exhibited higher performance accuracy in subjects $S_2$, $S_7$, $S_{13}$, and $S_{15}$. For subject $S_{14}$, the $\{r_{right}=+10, r_{wrong}=-30\}$ setting was more advantageous.

Although a smaller magnitude of $r_{wrong}$ relative to $r_{right}$ appears beneficial, a higher $r_{wrong}$ to $r_{right}$ ratio does not necessarily equate to improved accuracy. As demonstrated in Table~\ref{table: r_wrong vary}, performance accuracy diminishes with an increasing ratio, identifying the optimal ratio as twice the magnitude of $r_{wrong}$ to $r_{right}$. Furthermore, comparing different of reward configurations with equivalent magnitude ratios, such as $\{r_{right}=+5, r_{wrong}=-10\}$ and $\{r_{right}=+10, r_{wrong}=-20\}$, reveal subtle differences are noted in average performance accuracy and classification time. The former configuration outperforms in both average accuracy and time efficiency for classification. 

According to Table~\ref{table: r_wrong vary}, the classification time escalates with the increases in $r_{wrong}$ magnitude, where average times of 2.91, 3.41, 4.03, and 4.66 seconds were recorded for $r_{wrong}$ values of $-10$, $-20$, $-30$, and $-40$, respectively. This trend suggests that as the penalty for incorrect classification outweighs the reward for correct answers, the agents proceed with increased caution, hence extending the classification time.

\subsection{Effects of Episode Length Variation and Optimisation on Classification Performance}

In this study, we examined the mean accuracy, F1 score, and mean classification time across various episode lengths ($H$), including 10, 20, 30, and 40, as presented in Table~\ref{table: episode_length study}. We observed that both accuracy and F1 scores increased with extension of the episode horizon extends. Conversely, classification time per point increased with longer episode lengths. These finding suggests that larger episode lengths contribute to improvements in accuracy and F1 scores.

\begin{table}[ht]
\caption{Impact of Varying Episode Lengths ($H$) Values on Accuracy, F1 Score and Classification Time}
\label{table: episode_length study}
\centering
\setlength{\tabcolsep}{1.0pt}
\begin{tabular}{m{1.2cm}<{\centering}  m{2.3cm}<{\centering} m{2.3cm}<{\centering} m{2.0cm}<{\centering}}

\hline
\hline
Horizon $(H)$ & Accuracy\newline(Mean $\pm$ Std) & F1 Score\newline(Mean $\pm$ Std) & Mean Classification Time \\
\hline
10 & 94.46\% $\pm$ 8.10\% & 94.42\% $\pm$ 8.15\% & 2.18 \\
20 & 95.14\% $\pm$ 7.14\% & 95.10\% $\pm$ 7.18\% & 3.76 \\
30 & 95.56\% $\pm$ 6.54\% & 95.53\% $\pm$ 5.53\% & 5.53 \\
40 & 95.82\% $\pm$ 6.16\% & 95.79\% $\pm$ 6.54\% & 6.54 \\

\hline
\hline

\end{tabular}
\end{table}

Table~\ref{table: optimal rewards and lengths} delineates the optimal configuration of reward for correct ($r_{right}$) and incorrect ($r_{wrong}$) decisions, and episode horizon ($H$) that achieves the highest accuracy and F1 score in the shortest classification time possible. In this optimal setting, the RL agent demonstrates superior performance, achieving an average accuracy of 96.40\%  and an average classification time of less than 25 milliseconds across all 20 subjects.

\begin{table}[ht]
\caption{Subject-wise Classification Accuracy and Time with Optimal Reward Settings and Episode Lengths}
\label{table: optimal rewards and lengths}
\centering
\setlength{\tabcolsep}{1.0pt}
\begin{tabular}{m{0.8cm}<{\centering}  m{1.2cm}<{\centering}  m{1.2cm}<{\centering} m{1.4cm}<{\centering} m{1.4cm}<{\centering} m{1.2cm}<{\centering}}

\hline
\hline
Subj & Mean Accuracy & Mean F1 Score & Mean Classification Time & ($r_{right}$, $r_{wrong}$) & Episode Horizon \\
\hline
$S_{1}$ & \multicolumn{1}{r}{100.00\%} & \multicolumn{1}{r}{100.00\%} &  1.45 & (20, -10) & 10  \\
$S_{2}$ & \multicolumn{1}{r}{98.65\%} & \multicolumn{1}{r}{98.62\%} & 2.93 & (20, -30) & 30  \\
$S_{3}$ & \multicolumn{1}{r}{100.00\%} & \multicolumn{1}{r}{100.00\%} & 1.13 & (20, -10) & 10  \\
$S_{4}$ & \multicolumn{1}{r}{100.00\%} & \multicolumn{1}{r}{100.00\%} & 1.32 & (20, -30) & 10  \\
$S_{5}$ & \multicolumn{1}{r}{90.21\%} & \multicolumn{1}{r}{90.05\%} & 4.85 & (10, -10) & 30  \\
$S_{6}$ & \multicolumn{1}{r}{92.06\%} & \multicolumn{1}{r}{92.06\%} & 4.25 & (5, -10) & 40  \\
$S_{7}$ & \multicolumn{1}{r}{92.33\%} & \multicolumn{1}{r}{92.29\%} & 9.94 & (10, -30) & 40  \\
$S_{8}$ & \multicolumn{1}{r}{100.00\%} & \multicolumn{1}{r}{100.00\%} & 1.23 & (10, -10) & 10  \\
$S_{9}$ & \multicolumn{1}{r}{100.00\%} & \multicolumn{1}{r}{100.00\%} & 1.27 & (20, -10) & 10  \\
$S_{10}$ & \multicolumn{1}{r}{100.00\%} & \multicolumn{1}{r}{100.00\%} & 1.17 & (20, -10) & 10  \\
$S_{11}$ & \multicolumn{1}{r}{100.00\%} & \multicolumn{1}{r}{100.00\%} & 1.11 & (20, -20) & 10  \\
$S_{12}$ & \multicolumn{1}{r}{100.00\%} & \multicolumn{1}{r}{100.00\%} & 1.19 & (20, -10) & 10  \\
$S_{13}$ & \multicolumn{1}{r}{93.29\%} & \multicolumn{1}{r}{93.27\%} & 5.95 & (10, -10) & 40  \\
$S_{14}$ & \multicolumn{1}{r}{93.70\%} & \multicolumn{1}{r}{93.69\%} & 4.17 & (5, -10) & 40  \\
$S_{15}$ & \multicolumn{1}{r}{85.48\%} & \multicolumn{1}{r}{85.43\%} & 7.51 & (10, -20) & 40  \\
$S_{16}$ & \multicolumn{1}{r}{100.00\%} & \multicolumn{1}{r}{100.00\%} & 1.25 & (20, -10) & 10  \\
$S_{17}$ & \multicolumn{1}{r}{100.00\%} & \multicolumn{1}{r}{100.00\%} & 1.28 & (20, -10) & 10  \\
$S_{18}$ & \multicolumn{1}{r}{100.00\%} & \multicolumn{1}{r}{100.00\%} & 1.27 & (10, -10) & 10  \\
$S_{19}$ & \multicolumn{1}{r}{82.33\%} & \multicolumn{1}{r}{82.20\%} & 9.69 & (20, -30) & 40  \\
$S_{20}$ & \multicolumn{1}{r}{100.00\%} & \multicolumn{1}{r}{100.00\%} & 1.15 & (20, -10) & 10  \\
\hline
Mean & \multicolumn{1}{r}{96.40\%} & \multicolumn{1}{r}{96.38\%} & 3.21 &  - & - \\
Std & \multicolumn{1}{r}{$\pm$ 5.47} & \multicolumn{1}{r}{$\pm$ 5.50} & - & - & - \\
\hline
\hline

\end{tabular}
\end{table}

Our analysis, as indicated in Table~\ref{table: optimal rewards and lengths} shows that the RL agent achieved accuracy exceeding 90.00\% for each subject, with the exceptions of $S_{15}$ and $S_{19}$ whose accuracies were 85.48\% and 82.33\%, respectively. Subjects such as $S_1$, $S_3$, $S_4$, $S_8$, $S_9$, $S_{10}$, $S_{11}$, $S_{12}$, $S_{16}$, $S_{17}$, $S_{18}$, and $S_{20}$, where the RL agent achieved perfect classification, had notably clearer EEG MI signals. For these subjects, the agent performed consistently well across most reward and episode horizon configurations. Classifications were achieved within an average of 2 time points, where an optimal episode horizon of 10 and a reward configuration where $r_{right}$ significantly exceeded $r_{wrong}$ were conducive to faster classification decisions.

Particularly noteworthy was the performance of EEG\_RL-Net on subject $S_{13}$, where the RL agent achieved a classification accuracy of 93.29\%. This represented an exceptional improvement by 48.79\% over EEG\_GLT-Net with $m_{g\_GLT}$, the current state-of-the-art EEG MI time point classification method. The classification for $S_{13}$ took 6 time points on average, possibly reflecting the only subtle distinctions between EEG MI tasks for this subject.

\section{Conclusion}

Our study introduces EEG\_RL-Net, an innovative approach for the real-time classification of EEG-based motor imagery (MI) signals utilising reinforcement learning (RL) techniques. Building on the foundation of EEG\_GLT-Net's EEG\_GCN block and optimising computational efficiency with an adjacency matrix density of just 13.39\%, EEG\_RL-Net not only achieves accurate classification of EEG MI signals but also identifies signals that are unsuitable for classification. Remarkably, it achieved 100.00\% classification accuracy for 12 out of 20 subjects within less than 12.5 milliseconds. For challenging subjects ($S_{13}$ and $S_{19}$ in this study), where previous state-of-the-art methods such as EEG\_GLT-Net could classify with accuracies of only 44.50\% and 41.41\% respectively, EEG\_RL-Net achieved unprecedented improvement in performance, reaching classification accuracies of 93.29\% and 82.33\% in less than 62.5 milliseconds. These results underscore the robustness and efficacy of EEG\_RL-Net in enhancing classification rates, filling a gap for subjects previously deemed difficult by existing classification methods. In future work, we will further explore the integration of the optimal adjacency matrix $m_{g\_GLT}$ for advanced graph feature extraction in the EEG\_GCN block, aiming to unlock even greater improvements in the classification capabilities of our EEG\_RL-Net system.



\end{document}